\documentstyle[12pt]{article}
\def\fnote#1#2{\begingroup\def\thefootnote{#1}\footnote{#2}\addtocounter
{footnote}{-1}\endgroup}

\def\inbar{\vrule height1.5ex width.4pt depth0pt}
\def\IB{\relax{\rm I\kern-.18em B}}
\def\IC{\relax\,\hbox{$\inbar\kern-.3em{\rm C}$}}
\def\ID{\relax{\rm I\kern-.18em D}}
\def\IE{\relax{\rm I\kern-.18em E}}
\def\IF{\relax{\rm I\kern-.18em F}}
\def\IG{\relax\,\hbox{$\inbar\kern-.3em{\rm G}$}}
\def\IH{\relax{\rm I\kern-.18em H}}
\def\II{\relax{\rm I\kern-.18em I}}
\def\IK{\relax{\rm I\kern-.18em K}}
\def\IL{\relax{\rm I\kern-.18em L}}
\def\IM{\relax{\rm I\kern-.18em M}}
\def\IN{\relax{\rm I\kern-.18em N}}
\def\IO{\relax\,\hbox{$\inbar\kern-.3em{\rm O}$}}
\def\IP{\relax{\rm I\kern-.18em P}}
\def\IQ{\relax\,\hbox{$\inbar\kern-.3em{\rm Q}$}}
\def\IR{\relax{\rm I\kern-.18em R}}
\def\ZZ{\relax{\sf Z\kern-.4em Z}}
\def\beq{\begin{equation}}
\def\eeq{\end{equation}}
\def\bea{\begin{eqnarray}}
\def\eea{\end{eqnarray}}
\def\lleq#1{\label{#1}\eeq}
\def\llea#1{\label{#1}\eea}
\let\nn=\nonumber
\def\tabroom{\hbox to0pt{\phantom{\huge A}\hss}}
\def\notin{\ \hbox{{$\in$}\kern-.51em\hbox{/}}}

\def\a{\alpha}

  \def\cC{{\cal C}}

 \def\cS{{\cal S}}

\def\lra{\longrightarrow}
\def\lolra{\longleftrightarrow}

\def\hbar{\bar h}

 \def\II{{\bf II}} 
 
\def\fth{{{\rm F}_{12}}} 
\def\mth{{{\rm M}_{11}}}
\def\kthree{{\rm K3}} 
\def\twoto{{{\rm T}^2}} 
\def\fourto{{{\rm T}^4}}  
\def\twoa{{\rm IIA}} 
\def\het{{\rm Het}} 
\def\cyfour{{{\rm CY}_4}}
\def\cythree{{{\rm CY}_3}}

\begin{document}
\baselineskip=17pt
\parskip=.1truein
\parindent=0pt

\hfill {hep-th/9606148}
\vskip -.1truein
\hfill {BONN--TH--96--06}
\vskip -.1truein

\vskip .8truein

\centerline{\Large {\bf F--Theory on Calabi-Yau Fourfolds}}

\vskip .6truein
\centerline{\sc I.Brunner$^1$
                \fnote{\diamond}{Email: brunner@qft1.physik.hu-berlin.de }
                and
                R.Schimmrigk$^2$ 
              \fnote{\dagger}{Email: netah@avzw02.physik.uni-bonn.de}
            }

\vskip .5truein
\centerline{\it $^1$Institut f\"ur Physik, Humboldt Universit\"at, 
               10115 Berlin}
\vskip .01truein
\vskip .1truein
\centerline{\it $^2$Physikalisches Institut, Universit\"at Bonn, 
           53115 Bonn }
\vskip .01truein

\vskip 1.1truein
\centerline{\bf Abstract}

\noindent
We discuss some aspects of F-theory in four dimensions on 
elliptically fibered 
Calabi-Yau fourfolds which are Calabi-Yau threefold fibrations. 
A particularly simple class of such manifolds emerges for 
fourfolds in which the generic Calabi-Yau threefold fiber 
is itself an elliptic fibration and is K3 fibered. Duality 
between F-theory compactified on Calabi-Yau fourfolds and heterotic 
strings on Calabi-Yau threefolds puts constraints on the cohomology 
of the fourfold. By computing the Hodge diamond of Calabi-Yau 
fourfolds we provide first numerical evidence for F-theory 
dualities in four dimensions.

\renewcommand\thepage{}
\newpage

\baselineskip=17pt
\parskip=.1truein
\parindent=20pt
\pagenumbering{arabic}

\noindent
{\bf 1. Introduction}

\noindent
F-theory has proven to be a useful framework for many of the 
string dualities \cite{ht95,ew9503,kv95,duals}   
which have been discussed in the last two years. 
This fact indicates that 
F-theory \cite{cv96,mv96} 
(together with M-theory \cite{ew9503,js95,mth}) 
might lead to a higher dimensional embedding 
of various types of string theories. The emphasis of recent papers 
\cite{cv96,mv96,fthe} has been mostly on compactification 
of F-theory down to D=6 dimensions on Calabi-Yau 
threefolds\fnote{1}{F-theory on K3$\times $K3 has been
       considered in \cite{svw96}.}. 

In the present paper we discuss F-theory in four dimensions compactified on 
Calabi-Yau fourfolds. One of our tools is the generalization of the 
twist map of \cite{hs95}. This map provides an explicit construction 
of K3-fibered Calabi-Yau threefolds by starting from a 
specific K3 surfaces with 
an automorphism and an associated
 higher genus Riemann surface. It shows to what extent 
the Heterotic/Type II duality in D=4 can be traced   
to string/string duality in D=6. Furthermore it 
 isolates the additional structure 
of the fibration which is introduced by the twist of the fibration and 
is responsible for the dual type II image of the heterotic gauge structure. 
In Section 3 we generalize the twist map to construct
$(n+1)$-dimensional Calabi-Yau hypersurfaces which are fibered in 
terms of Calabi-Yau $n$-folds. We then use this map to derive and 
check consequences of F-theory duality in D$=$4.  
  
Starting from Vafa's duality conjectures in D=6,8 we 
apply the twist map  to Calabi-Yau threefolds to construct fourfolds 
whose generic fiber is the prescribed threefold. 
For F-theory one assumes that the Calabi-Yau space is an 
elliptic fibration. For such manifolds the expectation is that 
the twist map is concrete enough to allow for tests of the 
resulting D=4 conjecture  
relating\fnote{2}{We abbreviate Calabi-Yau $n$-folds by CY$_n$.}   
\beq
\fth(\cyfour) \lolra \het(\cythree).
\eeq
This conjecture thus provides a duality relation which involves 
a class of theories which are of phenomenological interest. 

One way to generate a class of CY$_3$-fibered Calabi-Yau fourfolds 
which are also elliptically fibered is by considering CY$_3$-fibrations 
for which the generic fiber is itself a K3-fibered threefold for which 
the K3 in turn is elliptic. For such manifolds the gauge structure 
results of Heterotic/Type II duality predicts the dimension of
cohomology groups of the Calabi-Yau fourfolds. This prediction 
can be tested. 
We take the first steps for such a check by computing the Hodge 
numbers for a variety of examples of Calabi-Yau fourfolds. The 
resolution structure of fourfolds is quite different from that 
of Calabi-Yau threefolds. We illustrate this difference by 
computing the cohomology of fourfolds
for a number of different fibration types. 

\vskip .2truein
\noindent
{\bf 2. F-Theory Dualities in Various Dimensions}

\noindent
In this section we work our way down from F-theory in D$=$8 to D$=$4 
dimensions. 

\noindent
{\bf 2.1 F-Theory in D=8}

\noindent 
It was argued in \cite{cv96} that F-theory in D=8 compactified on 
an elliptic K3 surface is dual to the heterotic string on T$^2$ 
\beq
\fth(\kthree) \lolra \het(\twoto),   
\eeq
the fibration of K3 being 
described by T$^2 \lra $K3 $\lra \IP_1$, where 
$\IP_1$ denotes the base of the fibration whose typical fiber is the torus 
T$^2$. 
Compactifying F-theory further on a torus leads to 
\beq
\fth(\kthree \times \twoto) \lolra \mth(\kthree \times S^1) \lolra 
\twoa(\kthree) \lolra \het(\fourto). 
\eeq

\noindent 
{\bf 2.2 F-theory in D=6}

\noindent 
In order to be able to push the D=8 duality of 2.1 
down to D=6 by lifting it from 
K3 surfaces to Calabi-Yau threefolds one considers elliptically 
fibered CY$_3$s. A simple class of this type are threefolds which are 
K3 fibrations for which the generic fiber in turn is 
elliptically fibered. These spaces are simultaneous of the type  
T$^2 \lra $CY$_3 \lra $B for some surface B, and of the type 
K3 $\lra $CY$_3 \lra \IP_1$. For such manifold one might expect 
\cite{cv96} to obtain the duality 
\beq
\fth({\rm CY}_3) \lolra \het(\kthree)  
\eeq
with the resulting chain of relations 
\beq
\fth(\cythree \times \twoto) \lolra \mth(\cythree \times S^1) \lolra 
\twoa(\cythree) \lolra \het(\kthree \times \twoto).      
\eeq

\noindent
{\bf 2.3 F-Theory in D=4} 

\noindent 
Our focus in the present paper is on compactifying F-theory down to 
four dimensions by considering Calabi-Yau fourfolds. 
To simplify the situation as much as possible we  
focus on fourfolds which are CY$_3$--fibered 
CY$_3 \lra $CY$_4 \lra \IP_1$ such that the threefolds defining 
the generic smooth fiber are in turn elliptically fibered 
K3-fibration. 
For such manifolds it is natural to expect the duality 
\beq
\fth(\cyfour) \lolra \het(\cythree) 
\eeq
and 
\beq
\fth(\cyfour \times \twoto) \lolra  \mth(\cyfour \times S^1) 
\lolra \twoa(\cyfour) \lolra \het(\cythree \times \twoto).
\eeq

\vskip .2truein 
\noindent 
{\bf 3. The Twist Map in Arbitrary Dimensions}

\noindent 
{\bf 3.1 The Twist Map}

\noindent 
In order to understand the way lower dimensional dualities can 
be inherited from the higher dimensional ones, and 
in particular to see to what extent this is possible at all, it 
is useful to have a tool which constructs  the necessary fibrations 
explicitly. Our way to do this employs 
the generalization of the orbifold construction of 
\cite{hs95} to arbitrary dimensions. In the following we 
will call this generalized 
map the twist map. Our starting point is a Calabi-Yau $n$--fold 
with an automorphism group $\ZZ_{\ell}$ whose action we denote 
by ${\bf m}_{\ell}$. Furthermore we choose
a curve $\cC_g$ of genus $g=(\ell -1)^2$ 
with projection $\pi_{\ell}: \cC_g \lra \IP_1$.  
The twist map then fibers Calabi-Yau $n$-folds into 
Calabi-Yau $(n+1)$-folds   
\beq
\cC_g \times {\rm CY}_n {\Big /}\ZZ_{\ell} \ni 
            \pi_{\ell}\times {\bf m}_{\ell}   
        ~~\lra ~~ {\rm CY}_{n+1}.
\eeq

For the class of weighted hypersurfaces 
\beq
\IP_{(k_0,k_1,...,k_{n+1})}[k] \ni 
\{y_0^{k/k_0}+p(y_1,...,y_{n+1})=0\},   
\lleq{hypsufib}
with $\ell = k/k_0 \in \IN$ and $k=\sum_{i=0}^{n+1}k_i$,   
the cyclic action can be defined as 
\beq
\ZZ_{\ell}\ni {\bf m}_{\ell}:~~~
  (y_0,y_1,...,y_{n+1}) \lra (\a y_0,y_1,...,y_{n+1}),  
\eeq
where $\a$ is the $\ell^{th}$ root of unity. An algebraic 
representation of the curve $\cC_g $ is provided by 
\beq
\IP_{(2,1,1)}[2\ell] \ni \{x_0^{\ell} 
             -\left(x_1^{2\ell}+x_2^{2\ell}\right) =0\}
\eeq
with action $x_0 \mapsto \a x_0$ and the remaining coordinates 
are invariant.
The twist map in this weighted context takes the form     
\beq
\IP_{(2,1,1)}[2\ell] \times \IP_{(k_0,k_1,...,k_{n+1})}[k]
{\Big /}\ZZ_{\ell} \lra \IP_{(k_0,k_0,2k_1,...,2k_{n+1})}[2k] 
\eeq
and is defined as 
\beq
((x_0,x_1,x_2),(y_0,y_1,...,y_{n+1})) \lra 
    \left(x_1 \sqrt{\frac{y_0}{x_0}},  
          x_2 \sqrt{\frac{y_0}{x_0}}, y_1,...,y_{n+1}\right) 
\lleq{twisthyp}

It is clear from the definition (\ref{twisthyp}) that the
twist map for hypersurfaces introduces additional singularites
on the fibered $(n+1)$--fold. In the simplest case, and the
case of interest in the present context this additional
singular set is the $\ZZ_2$ singular $(n-1)$-fold
$\IP_{(k_1,...,k_{n+1})}[k]$.
We will now focus on the special cases of $n=2$ and $n=3$ 
corresponding to the construction of weighted hypersurface 
fibrations of threefolds and fourfolds.

\noindent 
{\bf 3.2 Construction of Fibered CY-Threefolds}

\noindent 
For $n=2$ one finds that the map introduces the $\ZZ_2$-singular 
curve $C=\IP_{(k_1,k_2,k_3)}[k]$ which, on the threefold is 
in turn singular in general. When the resulting threefold
 fibrations are used in 
the context of Heterotic/Type II duality it is this additional 
curve which drastically changes the heterotic gauge structure 
one would expect if one were to focus solely on the K3 
fiber. The reason for this is that in the process of pushing 
down the D=6 Heterotic/Type II duality to four dimensions 
\beq
 \matrix{ \twoa(\kthree) &\lolra  &\het(\fourto) \cr 
             \downarrow  &      &\downarrow  \cr  
          \twoa(\cythree) &\lolra &\het(\kthree \times \twoto) \cr}
\eeq
the twist introduces branchings of the Dynkin resolution diagram 
of the K3 surface by glueing together the various disconnected 
resolution diagrams of the surface. Thus it is this twist which in 
addition to the 
K3 singularity structure determines the gauge group.  

\noindent
{\bf Example I:}
As an example consider K3 Fermat type 
surface in $\IP_{(1,2,6,9)}[18]$ with a
$\ZZ_{18}$ automorphism, the associated curve 
being $\IP_{(2,1,1)}[36]$.
The resulting K3-fibered threefold $\IP_{(1,1,4,12,18)}[36]$
has Hodge numbers $(h^{(1,1)},h^{(2,1)})=(7,271)$.
The heterotic gauge structure of this Calabi--Yau manifold
is determined by the curve $C=\IP_{(2,6,9)}[18]$ which glues
together the three $\ZZ_4$--points, whose resolution lead to
a total of 3 (1,1)--forms, and the $\ZZ_6$-point, whose resolution 
leads to 2 additional (1,1)-forms. Together with the K\"ahler
form of the ambient space these modes provide $h^{(1,1)}=7$.
The intersection matrix of the resolution is precisely given by
the Cartan matrix of the group SO(8)$\times $U(1)$^2$.

\noindent
{\bf Example II:}
We start with the K3 Fermat surface 
$\IP_{(1,6,14,21)}[42]$. $K$ has an automorphism 
group $\ZZ_{42}$ and we choose the curve as 
$\IP_{(2,1,1)}[84]$.
The image of the twist map is $\IP_{(1,1,12,28,42)}[84]$.
On the curve $C=\IP_{(6,14,21)}[42]$ one finds a $\ZZ_2$, a $\ZZ_3$
and a $\ZZ_7$ fixed point, leading to 1, 2 and 6 new curves,
respectively.
Hence we have $h^{1,1}=11$.
The resolution diagram is given by E$_8\times $U(1)$^2$
hence we see that the heterotic dual should be determined by Higgsing
the first E$_8$ completely while retaining the second E$_8$.
We also see that we should not fix the radii of the torus at some
particular symmetric point but instead embed the full gauge bundle
structure into the E$_8$.

\noindent 
{\bf Example III:} 
Our final threefold is based on the
 K3 surface $\IP_{(1,1,2,2)}[6]$ with a 
$\ZZ_6$ automorphism and the corresponding curve 
$\IP_{(2,1,1)}[12]$.
The resulting K3-fibration $\IP_{(1,1,2,4,4)}[12]$
has Hodge numbers $(h^{(1,1)},h^{(2,1)})=(5,101)$ and 
admits a conifold transition to a codimension two
Calabi--Yau manifold \cite{ls95}.
The heterotic gauge structure of this Calabi--Yau manifold
is determined by the curve $C=\IP_{(1,2,2)}[6] \sim \IP_2[3]$ 
which glues together the three $\ZZ_2$--points whose resolution 
leads to a total of 3 (1,1)--forms. Together with the K\"ahler
form of the ambient space we recover $h^{(1,1)}=5$.
The intersection matrix of the resolution is precisely given by
the Cartan matrix of the group SO(8) and we see that
in the heterotic dual we need to take the torus at the SU(3) point
in the moduli space and break this SU(3) by embedding the K3
gauge bundle structure groups appropriately. More details for this 
manifold, first discussed in this context in \cite{kv95}, can 
be found in \cite{hs95}.

\noindent
{\bf 3.3 Construction of Fibered CY-Fourfolds}

\noindent
When pushing down the duality 
\beq
\matrix{\twoa(\cythree) &\lolra  &\het(\kthree \times \twoto)\cr 
          \downarrow    &        &\downarrow \cr 
         \twoa(\cyfour) &\lolra  &\het(\cythree \times \twoto)\cr 
        }
\eeq
the singular curve on the generic Calabi-Yau 
fiber is embedded into the
$\ZZ_2$-singular surface $\IP_{(k_1,k_2,k_3,k_4)}[k]$.
In particular for the two K3 fibrations discussed above the 
only singularities that appear on the resulting fourfolds 
lie on the $\ZZ_2$-singular curves. 

Particularly simple classes of Calabi-Yau fourfolds, which are 
of interest in the context of duality, can be constructed 
by applying the twist map to the sequences of CY-threefolds 
discussed in \cite{mv96} and \cite{panda}. These result in 
the classes of fourfolds 
\bea
{\rm CY}_4^1(n)&:=&\IP_{(1,1,2,4n,8n+8,12n+12)}[24(n+1)] \nn \\
{\rm CY}_4^2(n)&:=&\IP_{(1,1,2,4n,4n+8,8n+12)}[8(2n+3))] \nn \\
{\rm CY}_4^3(n)&:=&\IP_{(1,1,2,4n,4n+8,4n+12)}[12(n+2)].  
\llea{cy4seqs}

To be concrete consider the images of the twist map of the three  
examples of threefolds discussed in Section 3.2. The first two 
of these lead to fourfolds 
in the first sequence of (\ref{cy4seqs}) 
 whereas the last example lives in neither 
of these classes. Now we know from the discussion above that 
the gauge group determined by the theory  
$\twoa(\IP_{(1,1,4,12,18)}[36])$ theory is
SO(8) $\times $U(1)$^4$. Pushing down this theory to
a fourdimensional F-theory via the twist map thus leads to the
prediction that the second cohomology group of the
fourfold $\IP_{(1,1,2,8,24,36)}[72]$, since it measures
the rank of the gauge group, should be 8-dimensional.
Similarly the Calabi-Yau threefold $\IP_{(1,1,12,28,42)}[84]$ leads to
the prediction that the rank of the second cohomology group of the
fourfold $\IP_{(1,1,2,24,56,84)}[168]$ should be 12
whereas the $\twoa(\IP_{(1,1,2,4,4)}[12])$ theory 
leads to the expectation that the second cohomology group of the 
fourfold $\IP_{(1,1,2,4,8,8)}[24]$
is 6-dimensional. 
It remains to compute the Hodge numbers of these spaces.

\vskip .2truein
\noindent 
{\bf 4. Calabi-Yau Fourfolds}

\noindent 
In this section we check the predictions of the previous 
discussion. The examples we focus on for this purpose are mostly 
contained in the class defined by the first sequence of hypersurface 
fourfolds described in (\ref{cy4seqs}), which are fibered as  
\beq
\matrix{ \IP_{(1,1,n,2n+4,3n+6)}[6(n+2)] &\lra 
              & \IP_{(1,1,2,2n,4n+8,6n+12)}[12(n+2)] \cr
              &        & \downarrow \cr
              &        & \IP_1\cr}. 
\eeq

Before coming to  the computation of the Hodge number for such 
fibrations, however, it should be noted that the `cohomology-behavior' 
of fourfolds is quite different from the behavior of threefolds.
In contrast to threefold hypersurfaces, for which the simplest member, 
$\IP_4[5]$ already leads to a representative cohomology Hodge 
diamond, this is not the case for Calabi-Yau fourfold hypersurfaces.
The fourfold analog of the threefold quintic, the smooth sextic 
fourfold, already illustrates this: 
for $\IP_5[6]$ the combined application of Lefshetz' hyperplane 
theorem, and counting complex deformations as well as computing 
the Euler number with the adjunction formula leads to 
the Hodge half-diamond  

\begin{scriptsize}
\begin{center}
\begin{tabular}{c c c c c c c c c}
     &   &     &   &1     &   &     &   &   \tabroom \\
     &   &     &0  &      &0  &     &   &   \tabroom \\ 
     &   &0    &   &1     &   &0    &    &   \tabroom \\ 
     &0  &     &0  &      &0  &     &0   &   \tabroom \\ 
   1 &   &426  &   &1752  &   &426  &    &1   \tabroom \\ 
 
\end{tabular}
\end{center}
\end{scriptsize}
 
\noindent
resulting in the Euler number $\chi_4=2610$. 
The point here is that the third cohomology group 
vanishes $b_3=0$, a fact that does not hold for general 4folds 
as we will see below. 

The methods just mentioned do not suffice for the computation 
of more general quasismooth Calabi-Yau fourfold hypersurfaces 
because one has to resolve the orbifold singularities. It turns 
out that the resolution of fourfolds is quite different from 
the resolution of threefolds. The resolution of both types of 
singularities, points \cite{ry87} as well as 
curves \cite{res} has been discussed in some detail for threefolds 
and differs markedly from the situation for fourfolds.
Furthermore in Calabi-Yau fourfolds we encounter the situation 
where we have to resolve surfaces. 

For any weighted Calabi-Yau fourfold one can use a combination 
of Cherning and resolution to compute the Euler number as 
\beq
\chi_4 = \int_{\cyfour } c_4 -\sum_i \frac{\chi(S_i)}{n_i} 
              + \sum_i n_i \chi(S_i),   
\eeq
where the $S_i$ are the $\ZZ_{n_i}$ singular sets of the manifold.
 
For fourfolds which are $\cythree$-fibered there is a quicker 
and independent way to do this computation by using the fibration 
formula developed in \cite{hs95}.  Given a fibration whose generic 
smooth fiber is a Calabi-Yau threefold 
which degenerates over a finite number of points $N_s$ of the base 
$\IP_1$ into a cone over a surfaces $\cS$ the Euler number follows.
For the class of hypersurfaces (\ref{hypsufib}) the formula becomes 
\beq
\chi(\cyfour) =(2-N_s)\chi(\cythree) + N_s(\chi(\cS)+k_0).
\lleq{eul4fib}
We postpone the detailed description of the geometric resolution 
to a more complete treatment and present here simply the 
results of our computations for the relevant manifolds.

We find for the Hodge half-diamond of the fourfold 
$\IP_{(1,1,2,8,24,36)}[72]$, the image under the twist map of 
our first example in Section 3.2,  

\begin{scriptsize}
\begin{center}   
\begin{tabular}{c c c c c c c c c}
     &   &     &   &1     &   &     &   &   \tabroom \\
     &   &     &0  &      &0  &     &   &   \tabroom \\
     &   &0    &   &8     &   &0    &    &   \tabroom \\
     &0  &     &0  &      &0  &     &0   &   \tabroom \\
   1 &   &6,528  &   &26,188  &  &6,528  &    &1   \tabroom \\

\end{tabular}
\end{center}
\end{scriptsize}

\noindent 
with Euler number $\chi_4=39,264$. The latter is in 
agreement with the computation via the fibration formula 
(\ref{eul4fib}). To see this it is sufficient to note that 
the generic smooth fiber degenerates over 72 points into the 
a surface $\cS$ of Euler number $\chi(\cS)=31$. 
For the Hodge diamond of the second example 
$\IP_{(1,1,2,24,56,84)}[168]$ we find 

\begin{scriptsize}
\begin{center}   
\begin{tabular}{c c c c c c c c c}
     &   &     &   &1     &   &     &   &   \tabroom \\
     &   &     &0  &      &0  &     &   &   \tabroom \\
     &   &0    &   &12     &   &0    &    &   \tabroom \\
     &0  &     &0  &      &0  &     &0   &   \tabroom \\
   1 &   &27,548  &   &110,284  &   &27,548  &    &1   \tabroom \\

\end{tabular}
\end{center}
\end{scriptsize}

\noindent 
with Euler number $\chi_4= 165,408$. We may check this   
with Cherning. To do so we need first to enumerate the 
singularities of the hypersurfaces, leading to 
  \bea
  \ZZ_2&:&S= \IP_{(1,12,28,42)}[84] \nn \\
  \ZZ_4&:&C= \IP_{(6,14,21)}[42] \nn \\
  \ZZ_8&:&\IP_{(3,7)}[21]=~1~pt \nn \\
  \ZZ_{12}&:&\IP_{(2,7)}[14]=~1~pt \nn \\
  \ZZ_{28}&:&\IP_{(2,3)}[6] =~1~pt, 
 \eea
and compute the Euler numbers of the singular surface and the 
curve. With the fourth Chern class 
$ c_4=222,223,000~h^4$ we then find 
     \bea
     \chi &=&\frac{27,777,875}{168}
       - \frac{1}{2}\left(\frac{1091}{84}+\frac{1}{84}\right)
       + 2\left(\frac{1091}{84}+\frac{1}{84}\right)
       -\frac{1}{4}\left(-\frac{1}{42}-\frac{1}{2} -
                         \frac{1}{3}-\frac{1}{7}\right) \nn \\  
      & &~~ + 4\left(-\frac{1}{42}-\frac{1}{2} -
           \frac{1}{3}-\frac{1}{7}\right)
      -\frac{1}{8} + 8 -\frac{1}{12} + 12 - \frac{1}{28} + 28 
   = 165,408. 
\eea
in agreement with the Hodge diamond. 

Finally, for the Hodge numbers of the third example 
$\IP_{(1,1,2,24,56,84)}[168]$ we find

\begin{scriptsize}
\begin{center}
\begin{tabular}{c c c c c c c c c}
     &   &     &   &1     &   &     &   &   \tabroom \\
     &   &     &0  &      &0  &     &   &   \tabroom \\
     &   &0    &   &6     &   &0    &    &   \tabroom \\
     &0  &     &1  &      &1  &     &0   &   \tabroom \\
   1 &   &803  &   &3,278  &   &803  &    &1   \tabroom \\

\end{tabular}
\end{center}
\end{scriptsize}

\noindent 
with Euler number $\chi_4=4,896$, which again can be checked with  
either the fibration formula or Cherning.  

These numbers confirm the predictions of the analysis in the 
previous section. In Table 1 we list the results for 
a few other fourfolds. The emerging structure shows that 
the nonvanishing of $b_3$ is tied to the existence of singular 
sets of dimension one. 
It should be noted that the Euler number 
of all fibered manifolds 
considered here are divisible by 24, a necessary condition 
for anomaly cancellation\cite{vw95}.   

\begin{scriptsize}
\begin{center}
\begin{tabular}{| l l r r r r r l |}
\hline
No.  &Manifold        &$\chi$      &$h^{(1,1)}$  
   &$h^{(2,1)}$ &$h^{(3,1)}$ &$h^{(2,2)}$  &CY$_3$ Fibers,~~Gauge Group
                                                     \tabroom \\
\hline
1  &$\IP_5[6]$      &2,610  &1   &0  &426  &1,752   
                 & --  \tabroom \\
2   &$\IP_{(1,1,2,2,2,2)}[10]$   &2,160 &2   &0   &350  
        &1452    &$\IP_4[5]$     \tabroom  \\
3   &$\IP_{(1,1,2,2,6,6)}[18]$   &4,176  &5  &0  &683 
               &2,796   &$\IP_{(1,1,1,3,3)}[9]$\tabroom  \\
4   &$\IP_{(1,1,2,4,4,4)}[16]$  &2,688   &3  &3     
               &440 &1,810   &$\IP_{(1,1,2,2,2)}[8]$ \tabroom \\
5   &$\IP_{(1,1,2,4,4,12)}[24]$  &6,096 &3 &2   
           &1,007  &4,080   
              &$\IP_{(1,1,2,2,6)}[12]$,~U(1)$^3$  \tabroom \\
6   &$\IP_{(1,1,2,4,8,8)}[24]$ &4,896  &6 &1  
             &803  &3,278   
     &$\IP_{(1,1,2,4,4)}[12]$,~SO(8)$\times $U(1)$^2$
                      \tabroom \\
7   &$\IP_{(1,1,2,4,16,24)}[48]$ &23,328
     &4  &1 &3,876  &15,566  
     &$\IP_{(1,1,2,8,12)}[24]$,~U(1)$^4$ 
                  \tabroom \\
8   &$\IP_{(1,1,2,8,24,36)}[72]$ &39,264 
     &8  &0 &6,528  &26,188 
     &$\IP_{(1,1,4,12,18)}[36]$,~SO(8)$\times $U(1)$^4$
                  \tabroom \\  
9   &$\IP_{(1,1,2,24,56,84)}[168]$ &165,408
     &12   &0 &27,548 &110,284  
     &$\IP_{(1,1,12,28,42)}[84]$,~E$_8 \times $U(1)$^4$
                    \tabroom \\
\hline
\end{tabular}
\end{center}
\end{scriptsize}

\noindent  
{\bf Table}{~\it A short list for the cohomology of Calabi-Yau 
      fourfolds of different fibration type. We also the record 
     the known gauge groups.}

\vskip .2truein
\noindent 
{\bf 5. Conclusion}

\noindent 
We have shown that pushing down the duality relations from 
$\fth(\kthree)$ to $\fth(\cyfour)$ 

\begin{small}
\beq
\matrix{ \fth(\kthree \times \twoto) &\lolra &\mth(\kthree \times S^1) 
                                     &\lolra &\twoa(\kthree) 
                                     &\lolra &\het(\fourto)  \cr
         \downarrow                 &        &\downarrow 
                                    &        &\downarrow 
                                    &        &\downarrow \cr  
         \fth(\cythree \times \twoto) &\lolra &\mth(\cythree \times S^1)
                                &\lolra & \twoa(\cythree)
                                &\lolra &\het(\kthree \times \twoto)  \cr
         \downarrow                 &        &\downarrow
                                    &        &\downarrow
                                    &        &\downarrow \cr
         \fth(\cyfour \times \twoto) &\lolra &\mth(\cyfour \times S^1)
                               &\lolra & \twoa(\cyfour)
                                &\lolra &\het(\cythree \times \twoto)  \cr
       }
\eeq
\end{small}

\noindent
via the generalized twist map leads to predictions which can be 
confirmed. 

\vskip .2truein
\noindent
{\bf Acknowledgement}  

\noindent
It is a pleasure to thank Paul Green, Tristan H\"ubsch,  
Bruce Hunt and Dieter L\"ust for discussions. 
We would also 
like to thank the Erwin Schr\"odinger Institute, Vienna, and the 
Mathematics Institute, Oberwolfach, for hospitality.

\end{document}